\documentclass[12pt]{iopart}
\usepackage{hyperref,graphicx,enumitem,xcolor}
\usepackage{harvard}
\newcommand{\figFRAMEWORK}{
\begin{figure}
\hspace{2.45cm}
\includegraphics[width=13.15cm]{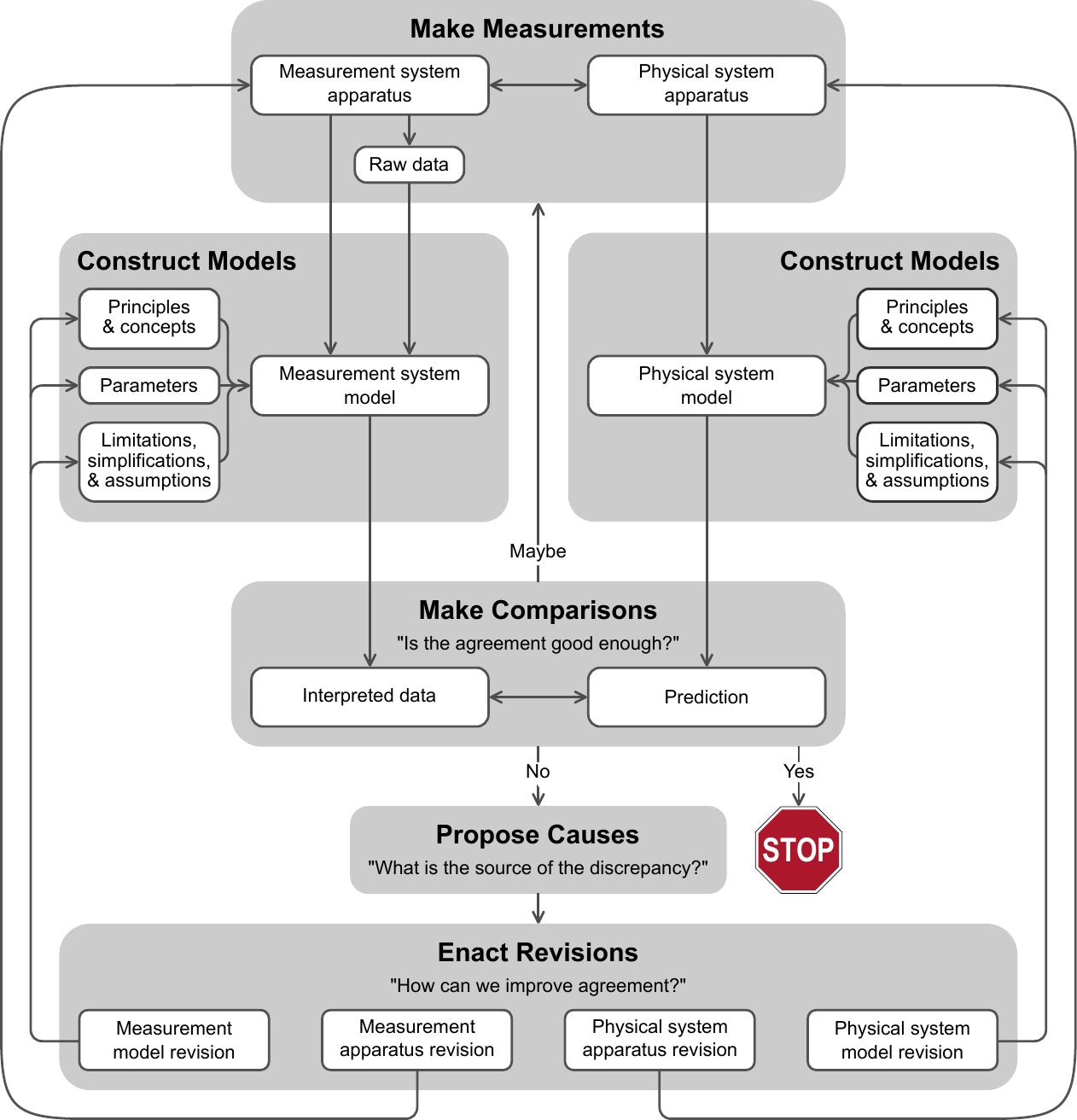}
\caption{\label{fig:framework}Modeling Framework for Experimental Physics. The version of the Framework presented here was previously published by Dounas-Frazer \etal (2018) and is adapted from a visualization originally developed by Zwickl \etal (2014).}
\end{figure}
}

\newcommand{\figMALUS}{
\begin{figure}
\hspace{2.45cm}
\includegraphics[width=13.15cm]{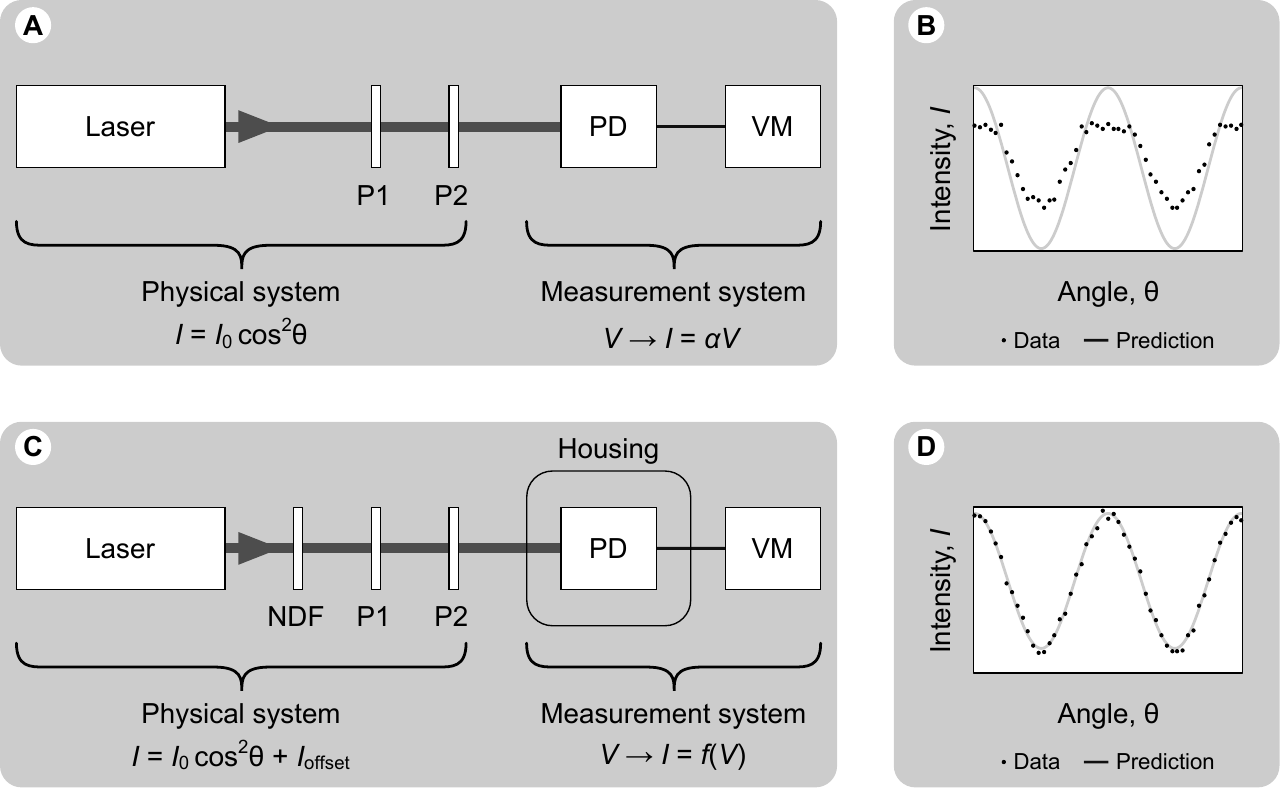}
\caption{\label{fig:malus}{Hypothetical Malus's Law experiment. (A) Initial setup. P1 and P2 are polarizers. PD is a photodetector. VM is a voltmeter. $I_0$ and $I$ are the intensities of laser light incident on P2 and PD. $\theta$ is the angle between axes of polarization of P1 and P2. $\alpha$ is related to the linear responsivity of the PD. (B) Disagreement between data and prediction. (C) Revised setup. NDF is a neutral density filter. $I_\mathrm{offset}$ is a modification to Malus's Law. $f(V)$ represents nonlinear calibration of PD output to optical power. (D) Improved agreement between data and prediction.
}}
\end{figure}
}

\begin{document}


\title[Modeling Framework for Experimental Physics]{The Modeling Framework for Experimental Physics: \\ Description, development, and applications}

\author{Dimitri R. Dounas-Frazer and H. J. Lewandowski}
\address{Department of Physics, University of Colorado Boulder, Boulder, CO, USA}
\address{JILA, National Institute of Standards and Technology and University of Colorado Boulder, Boulder, CO, USA}
\ead{dimitri.dounasfrazer@colorado.edu}
\ead{lewandoh@colorado.edu}

\vspace{10pt}
\begin{indented}
\item[]\today
\end{indented}

\begin{abstract}
The ability to construct, use, and revise models is a crucial experimental physics skill. Many existing frameworks describe modeling in science education at introductory levels. However, most have limited applicability to the context of upper-division physics lab courses or experimental physics. Here, we discuss the Modeling Framework for Experimental Physics, a theoretical framework tailored to labs and experimentation. A key feature of the Framework is recursive interaction between models and apparatus. Models are revised to account for new evidence produced by apparatus, and apparatus are revised to better align with the simplifying assumptions of models. Another key feature is the distinction between the physical phenomenon being investigated and the measurement equipment used to conduct the investigation. Models of physical systems facilitate explanation or prediction of phenomena, whereas models of measurement systems facilitate interpretation of data. We describe the Framework, provide a chronological history of its development, and summarize its applications to research and curricular design. Ultimately, we argue that the Modeling Framework is a theoretically sound and well-tested tool that is applicable to multiple physics domains and research purposes. In particular, it is useful for characterizing students' approaches to experimentation, designing or evaluating curricula for lab courses, and developing instruments to assess students' experimental modeling skills.
\end{abstract}

\submitto{\EJP}

\section{\label{sec:models}Models and modeling in physics education}
The importance of modeling in physics education has been recognized for over 30 years. \citeasnoun{Johnson-Laird1980}, \citeasnoun{Hestenes1987}, and \citeasnoun{Halloun1987} played key roles in the development of early theories of learning and curricular design. Today, students' development of mental models \cite{Brewe2018,Korhasan2016,Ozcan2015} and representational competence \cite{Fredlund2012,McPadden2017,Opfermann2017} continue to be active areas of research {in lecture or studio settings}. Meanwhile, undergraduate physics lab courses have been gaining increased attention over the last decade~\cite{Schumacher2007}. Consider, for example, multiple large survey studies of students' views about lab courses \cite{Coppens2016a,Deacon2011,Hanif2009} or experimental physics \cite{Wilcox2018}. Despite these trends, research on student engagement in modeling during lab courses is sparse, and there is a need for theoretical frameworks and empirical studies that focus on students' ability to model experimental systems.

Since the 1980s, conceptions of models originally rooted in cognitive science have been supplemented with philosophical arguments \cite{Gilbert2016a}. \citeasnoun{Knuuttila2011} argues that models are tools for generating knowledge. As tools, models consist of representations that are expressed externally via a material medium, e.g., ink and paper. Mental models, on the other hand, can be precursors to external representations, and are also part of the modeling process \cite{Gilbert2016a}. To facilitate knowledge generation, models support people to ask questions, recognize and explain patterns in data, and make judgements about those explanations \cite{Passmore2012}. {Models can also be used to communicate new knowledge or understanding to others \cite{Schwarz2009}}. However, tools do not spontaneously work on their own. Indeed, \citeasnoun{Giere2009} and \citeasnoun{Gouvea2017} make bids for focusing on the agents and ends of modeling: people use models to represent part of the world for some purpose.

In physics education, the purpose of modeling is often for students to describe, explain, or predict the behavior of physical phenomena~\cite{Etkina2006}. Students use evidence to generate models, and they use models to search for new evidence. {Such intertwining of evidence- and model-based reasoning supports students to make sense of phenomena} \cite{Russ2017}. \citeasnoun{Koponen2007} argues that modeling in physics is a bidirectional process through which models both inform, and are informed by, experimental apparatus. Apparatus are designed to isolate particular aspects of a model, and data are used to verify the model or identify its limitations. Koponen further claims that two models are at play during {experimentation}: one to formulate explanations or predictions about a physical system that is the target of investigation, and another to interpret the experimental data generated by the measurement equipment. Thus, in physics education, the practice of modeling should involve recursive interactions between apparatus, evidence, and models of both phenomena and equipment. {More broadly, we view such interactions as key features of experimental physics practice. Although many theoretical frameworks for modeling exist \cite{Brewe2008,Fuhrmann2018,Gilbert2016a,Passmore2012,Schwarz2009,Windschitl2008a}, few capture these key features}.

Here, we {discuss} the Modeling Framework for Experimental Physics, which was explicitly designed to characterize physicists' use of models when conducting experiments. In the next section, we describe the Framework using a Malus's Law experiment to explicate various modeling subtasks and processes. Then, in Secs.~\ref{sec:development} {and \ref{sec:validation}}, we provide an overview of the Framework's initial development and {its subsequent validation and use as a research tool}. Sec.~\ref{sec:applications} discusses applications to the design of undergraduate lab courses. Finally, in Sec.~\ref{sec:conclusion}, we argue that the Modeling Framework is a dual-purpose tool that is theoretically sound, well-tested, and {versatile.  By ``versatile," we mean that the Framework is applicable to multiple physics domains, it can be used to characterize student and instructor behaviors, and it can inform the design and evaluation of curricula for lab courses.}

\section{\label{sec:description}Description of the Modeling Framework for Experimental Physics}
A visual representation of the Modeling Framework for Experimental Physics is given in Fig.~\ref{fig:framework}. the Framework can be thought of as a flowchart that consists of five subtasks: making measurements, constructing models of equipment and phenomena, making comparisons between data and predictions, proposing causes for discrepancies, and enacting revisions to models and apparatus. Because modeling is a recursive process, the flowchart is cyclical. In the Framework, the goal is to achieve sufficiently good agreement between data and predictions, {consistent with other work on modeling in physics education \cite{Etkina2006}}. 

\figFRAMEWORK

{Below, we elaborate on each subtask of the Modeling Framework, drawing on examples from a Malus's Law experiment to explicate abstract concepts. This experiment involves shining laser light through two linear polarizers, one that is fixed and one that is free to rotate (the analyzer). After passing through both polarizers, the intensity of the transmitted laser light is proportional to $\cos^2\theta$, where $\theta$ is the angle between the axes of polarization of the two polarizers. Thus, by rotating the analyzer, students can vary the intensity of the transmitted light. A schematic of the experiment is provided in Fig.~\ref{fig:malus}.}

\figMALUS

\begin{enumerate}
\item \textbf{Make measurements.} The measurement equipment interacts with the physical system and produces raw data.\\
\emph{Malus's Law example.} The physical system apparatus includes a laser and two linear polarizing filters. The measurement equipment consists of a photodetector and {voltmeter}. Making measurements involves shining light through the polarizers onto the photodetector {and producing a voltage on the voltmeter. See Fig.~\ref{fig:malus}(A).}
\item \textbf{Construct models.} The physical and measurement system apparatus are associated with distinct models. Each model is created from the key features of the system, relevant physics principles and concepts, particular parameter values, and appropriate assumptions and simplifications that make the model tractable, while limiting its scope of applicability.\\
\emph{Malus's Law example.}
The simplest model of the physical system involves treating the laser light as a perfectly monochromatic plane wave with uniform intensity. Polarizers are similarly idealized as {having extinction ratios of exactly zero}. The simplest measurement system model treats the measurement equipment as a black box: the photodetector instantaneously converts incident laser light to an electrical signal that is sent to the voltmeter, which {produces} a voltage linearly proportional to the {power} of the light.
\item \textbf{Make comparisons.} The physical system model is used to generate predictions, and the measurement system model is used to analyze and interpret raw data. Interpreted data are compared to predictions, and domain-specific criteria are used to determine whether the level of agreement is good enough. {Applying accepted scientific criteria for determining what counts as ``good enough" is part of the modeling process \cite{Giere2009,Gouvea2017}}. If the agreement is good enough, then there is no need to revise models or apparatus. However, additional data are sometimes needed in order to properly judge the quality of agreement; in this case, more measurements {are required}. If there is a discrepancy between data and predictions, the experimenter may propose and enact changes to models or apparatus to resolve the discrepancy. In this sense, the ``Maybe" and ``No" pathways in Fig.~\ref{fig:framework} respectively correspond to efforts to reduce statistical and systematic sources of uncertainty.
\\
\emph{Malus's Law example.} The physical system model is used to predict Malus's Law: $I = I_0\cos^2\theta$, where $I_0$ and $I$ are the intensities of light {incident on the analyzer and photodetector, respectively}. The measurement system model is used to interpret raw data by converting voltage as a function of angle to normalized intensity as a function of angle. {Experimental data are compared to a cosine-squared curve; see Fig.~\ref{fig:malus}(B). Goodness-of-fit metrics can help determine quality of agreement.} Many such metrics have established {criteria for acceptable fits} among physicists.
\item \textbf{Propose causes.} When discrepancies between data and predictions must be resolved, {physicists} generate hypotheses about potential sources of {those discrepancies}. This modeling subtask can be challenging for people who are unfamiliar with nonideal behavior of equipment or assumptions that limit the predictive power of models \cite{Zwickl2015a,Dounas-Frazer2018arXiv}.\\
\emph{Malus's Law example.} {Common sources of discrepancy between data and prediction include nonzero background due to ambient light, polarizers with nonzero extinction ratios, or laser light with nonzero ellipticity. Discrepancies may also arise if the light is sufficiently powerful to saturate the photodetector or if changes in light power are faster than the slew rate limit of the detector.}
\item \textbf{Enact revisions.} The Modeling Framework describes four pathways for revision: one may revise the models or apparatus of the physical or measurement systems. These pathways are consistent with other conceptions of modeling as a process that informs the design and execution of experiments \cite{Gilbert2016a,Koponen2007,Russ2017}. Depending on which proposed causes are deemed most likely or easiest to implement, some revisions may be prioritized over others \cite{Stanley2017}. Once revisions are enacted, the modeling cycle repeats until data and predictions are brought into good enough agreement.\\
\emph{Malus's Law example.} Apparatus revisions include covering the photodetector to shield it from ambient light, or placing a neutral density filter in the optical path of the laser light to decrease its overall intensity. Model revisions include adding a fit variable to Malus's Law to represent offsets due to ambient light or ellipticity, or determining the nonlinear calibration of the photodetector for high light intensities. {See parts (C) and (D) of Fig.~\ref{fig:malus}.}
\end{enumerate}

\section{\label{sec:development}Development of the Modeling Framework}
The Modeling Framework (Fig.~\ref{fig:framework}) was originally developed in the context of the transformation of the Advanced Laboratory course at the University of Colorado Boulder (CU) \cite{Zwickl2013,Zwickl2014}. Those efforts began in late 2010 {and employed the following approach}: define learning goals, develop curriculum that align with those goals, and assess student learning. The Modeling Framework emerged from the first two phases of the transformation process. To identify consensus learning goals for the course, \citeasnoun{Zwickl2013} worked with 21 faculty members via individual interviews or group meetings. Learning goals were also informed by comparison of the Advanced Laboratory to other laboratory courses and a review of the physics education literature. Using this process, \citename{Zwickl2013} identified modeling as a major learning goal for the course. They further argued that modeling had three components: modeling the physical system, modeling the measurement system, and comparing data and predictions.

After identifying modeling as a learning goal for the Advanced Laboratory, \citeasnoun{Zwickl2014} developed laboratory activities designed to engage students in the practice of modeling during optics experiments {(Sec.~\ref{sec:applications})}. To guide their designs, they created the first iteration of the Modeling Framework. This iteration was informed by a review of the physics education literature and {Zwickl and Lewandowski's} own experience and expertise as experimental physicists. The first iteration differs from that presented in Fig.~\ref{fig:framework} in several ways, most of them aesthetic. Nevertheless, the main features have persisted across all versions of the Framework: measurement and physical systems are distinct, revisions include changes to both apparatus and models, and the need for revision is triggered by comparisons that fail to yield ``good enough" agreement between interpreted data and model predictions. Thus, from its inception, {the Modeling Framework aligned with philosophical conceptions of modeling, such as \possessivecite{Koponen2007} bidirectionality}.

\section{\label{sec:validation}{Validation and use of the Modeling Framework as a research tool}}

{After the Modeling Framework was {developed}, it was validated and used as a research tool in a series of four studies, three focused on students \cite{Dounas-Frazer2016a,Stanley2017,Zwickl2015a} and one on instructors \cite{Dounas-Frazer2018arXiv}. These studies demonstrate that the Modeling Framework is appropriate for characterizing modeling in upper-division lab contexts. They also resulted in small changes that improved the descriptive power of the Framework.}

\subsection{\label{sec:students}{Student approaches to modeling during optics and electronics activities}}

\citeasnoun{Zwickl2015a} conducted think-aloud interviews during which eight students verbalized their reasoning while testing the power output of a light-emitting diode. {One major contribution of this investigation was validation of the Modeling Framework's focus on measurement system models. During interviews, student engagement with the measurement apparatus (i.e., a photodetector and oscilloscope) included constructing models, revising models, interpreting output, and identifying limitations of the equipment.} \citename{Zwickl2015a} also described two common barriers to students' use or revision of models. First, {some students did not articulate crucial assumptions and corresponding limitations of their models, preventing them from making model-based refinements to their experiment}. Second, lack of familiarity with the concept of solid angle and the unit of steradian prevented some students from appropriately comparing their measurements to numerical values specified in the data sheet for the light-emitting diode. These barriers are consistent with other work that has demonstrated how students' prior knowledge impacts their ability to construct and evaluate models~\cite{Fortus2016,Ruppert2017,Stewart2005}.

{Next, \citeasnoun{Dounas-Frazer2016a} expanded on previous work by demonstrating that the Modeling Framework is applicable to a context other than optics, namely, electronics.} In that work, eight pairs of students from two institutions participated in think-aloud interviews that involved {troubleshooting} a malfunctioning electric circuit. Troubleshooting can be thought of as a type of modeling in which the physical circuit is revised in order to bring its performance into alignment with expectations informed by a model. \citename{Dounas-Frazer2016a} examined students' modeling behaviors during two key episodes in the troubleshooting process: isolating the source of malfunction to a particular  subsystem, and evaluating the performance of the repaired circuit. In the former episode, students engaged in constructing models and making comparisons more often than other subtasks; in the latter episode, making comparisons, proposing causes, and enacting revisions were the most common subtasks. Thus, not only did students engage in multiple modeling cycles throughout the troubleshooting process, different phases of troubleshooting corresponded to different {combinations of subtasks}. Because troubleshooting is a central feature of electronics lab courses \cite{Dounas-Frazer2017}, this study suggests that so, too, is modeling.

{Transitioning from clinical research settings to an actual classroom, \citeasnoun{Stanley2017} used the Modeling Framework to study student use of models in an Electronics Laboratory course specifically designed to engage students in modeling circuits \cite{Lewandowski2015}.} They analyzed 45 student lab notebooks for evidence of students' documented engagement in modeling subtasks. Three notebook entries were selected for analysis, each corresponding to a different activity: voltage divider circuit (high scaffolding), photometer circuit (medium scaffolding), and voltage-controlled electromagnet (low scaffolding). Student engagement in modeling tracked the level of scaffolding in the lab guide. Compared to open-ended prompts, explicit prompts resulted in more thorough engagement in modeling. In particular, most students did not enact revisions to their models or apparatus unless specifically asked to do so. Further, \citename{Stanley2017} found that students often neglected to make and document comparisons between expected and actual performance of circuits even when prompted, potentially due to lack of clarity about what constitutes ``good enough" agreement between data and predictions. Making comparisons and judging the quality of agreement is crucial for making informed decisions about whether and how to iteratively improve an experiment \cite{Holmes2015b}. Therefore, there is a clear need to better understand and support student reasoning about {which standards for ``good enough" agreement are appropriate in different experimental physics contexts (cf. \citename{Giere2009} 2009, \citename{Gouvea2017} 2017). In addition to identifying areas for curricular improvement and future study, \citename{Stanley2017} demonstrated that the Modeling Framework is useful for understanding student modeling in formal educational settings.}

\subsection{\label{sec:instructors}{Instructor perspectives on modeling in optics and electronics lab courses}}

Shortly after it was created, the Modeling Framework informed a set of nationally endorsed recommendations for undergraduate physics lab courses in the United States~\cite{AAPT2015}. However, until recently, research on modeling in upper-division physics labs focused almost exclusively on the perspectives and behaviors of instructors and students from CU \cite{Dounas-Frazer2016a,Stanley2017,Zwickl2014,Zwickl2015a}.

To get a better idea of whether and how modeling is taken up in lab courses across the United States, \citeasnoun{Dounas-Frazer2018arXiv} conducted interviews with {19 optics instructors and 16 electronics instructors} from 27 institutions. During these interviews, instructors described how various subtasks of the Modeling Framework aligned with their learning goals or activity design. Making measurements, constructing models, and comparing data to predictions were each identified as important learning goals by a majority of instructors in the study. Enacting revisions and proposing causes were less commonly identified as important. Limited class time was cited as a barrier to student revisions to experiments. Meanwhile, many instructors said that students are unfamiliar with the nonideal behavior of devices, and therefore struggle to propose causes for discrepancies between data and predictions. This finding is consistent with students' failure to recognize or articulate assumptions and limitations of models during experimental physics think-aloud interviews \cite{Zwickl2015a}.

Further, \citeasnoun{Dounas-Frazer2018arXiv} found that modeling is taken up differently in optics compared to electronics. For example, with respect to the subtask of making comparisons, optics instructors were more likely to describe engaging students in rigorous statistical analyses (e.g., fitting curves to data and evaluating the goodness of fit). In contrast, electronics instructors said that comparisons typically involved qualitative checks of circuit performance because building functional circuits was more important than achieving precise agreement between predicted and expected output voltages. Electronics instructors often framed particular modeling subtasks or the whole Modeling Framework as necessary aspects of troubleshooting, in alignment with students' approaches to repairing malfunctioning circuits during think-aloud interviews \cite{Dounas-Frazer2016a}. Overall, \citeasnoun{Dounas-Frazer2018arXiv} not only demonstrated the versatility of the Framework for describing the goals and activities of a national sample of lab courses, they also shed light on the different purposes of modeling across two experimental physics domains.

\subsection{{Empirically motivated changes to the Modeling Framework}}

Over the course of the investigations {described in Secs.~\ref{sec:students} and \ref{sec:instructors}}, the Modeling Framework itself was modified in order to better capture the experimental modeling process. While many changes were aesthetic, some represented shifts in understanding of what modeling entails. In the first visualization of the Framework , \citeasnoun{Zwickl2014} did not explicitly include ``limitations" as part of system models, even though they recognized that students should be able to articulate model limitations. After \citeasnoun{Zwickl2015a} observed that students' unarticulated assumptions and unrecognized model limitations were barriers to modeling, they revised the Framework  to include ``limitations, simplifications, and assumptions" as an explicit part of system models. The ``Maybe" pathway was added by \citeasnoun{Dounas-Frazer2018arXiv} based on experience observing students troubleshoot electric circuits \cite{Dounas-Frazer2016a}. When confronted with a disagreement between data and predictions, students sometimes collected additional data in order to be sure that the discrepancy was significant, rather than immediately trying to explain it. These and other empirically motivated changes to the Modeling Framework likely improve its usefulness as a {well-tested} tool for describing the experimental modeling process.

\section{\label{sec:applications}Applications of the Modeling Framework to curriculum design}
The Modeling Framework (Fig.~\ref{fig:framework}) is a dual-purpose tool that can both characterize people's reasoning about experimental systems and inform curriculum design for laboratory courses~\cite{Zwickl2014}. To date, we are aware of applications of the Framework  to the design of three undergraduate contexts: a fourth-year Advanced Laboratory~\cite{Zwickl2014}, a third-year Electronics Laboratory~\cite{Lewandowski2015}, and first-year introductory courses~\cite{Vonk2017}.

The first educational application of the Modeling Framework coincided with its development, as described in {Sec.~\ref{sec:development}}. After establishing that students' ability to model experimental systems was a learning goal for the CU Advanced Laboratory~\cite{Zwickl2013}, \citeasnoun{Zwickl2014} developed and used the Framework  to guide their transformation of that course. Soon after, \citeasnoun{Lewandowski2015} also transformed the CU Electronics Laboratory in order to meet similar learning goals. Both courses aimed to engage students in constructing models, making predictions and comparisons, and revising models and apparatus. Inspired by previous work on scaffolded inquiry \cite{Buck2008,Etkina2008,Hmelo-Silver2007,Kirschner2006}, lab guides explicitly prompted students to engage in these modeling subtasks. For example, the lab guide for an Advanced Laboratory polarization activity prompted students to use the Jones formalism to model the propagation of laser light through polarizing filters and wave plates, and to derive predictions like Malus's Law \cite{Zwickl2014}. To model the measurement system, students used manufacturer documentation to understand the operation and limitations of a photodetector, and to appropriately convert output voltage into measurements of optical power. Similarly, in the Electronics Laboratory, the lab guide for a voltage divider circuit prompted students to revise their circuit schematic and equation for the transfer function in order to include the input resistance of a digital multimeter. The level of explicit scaffolding in the lab guides faded over the course of the semester in order to provide students with more control over how they modeled their circuits \cite{Lewandowski2015}. In both courses, lab guides prompted students to reason about model limitations or revisions in order to explain or minimize systematic biases in their measurements. 

At the introductory level, \citeasnoun{Vonk2017} used the Modeling Framework to design model-making activities in introductory algebra- and calculus-based physics courses. Here ``model-making" refers to devising an experiment to determine the relationship between two variables (e.g., wavelength and frequency of a wave on a string), collecting and analyzing data, constructing a model to relate the variables, and using the model to make predictions. Instead of working with apparatus, students used Direct Measurement Videos (DMVs) to explore phenomena and collect data. DMVs are ``short high-quality videos that show a scientifically interesting event," {which are} analyzed using online tools like digital rulers and stopwatches \cite[p.4]{Vonk2017}. Because students did not use apparatus, they engaged in only some modeling subtasks: making measurements, constructing models of physical phenomena, and making comparisons. In a study of 116 students' performance on a model-making assessment, \citename{Vonk2017} showed that students who completed activities using {DMVs designed to bolster model-making skills} outperformed those who did not. This work suggests that the Modeling Framework has implications for a wide range of course formats and activity types beyond physics labs and apparatus-based activities.

\section{\label{sec:conclusion}Summary and ongoing work}
The Modeling Framework for Experimental Physics (Fig.~\ref{fig:framework}) was developed to describe the process of constructing, using, and revising models when conducting physics experiments. {It was informed by, and is consistent with, other theoretical conceptions of scientific modeling.} Multiple studies have demonstrated that there is a good empirical mapping between the Framework  and students' approaches to completing experimental physics tasks and instructors' learning goals and activity design in upper-division lab courses. Additionally, the Framework  has been used to inform the design and evaluation of lab courses and introductory algebra- and calculus-based physics courses. For these reasons, we argue that the Modeling Framework is a theoretically sound, well-tested, and versatile tool.

The multi-year process through which the Modeling Framework was developed has yielded some insights about barriers to engaging students in modeling during upper-division physics labs. {In these courses, students' prior knowledge about relevant physics concepts or model limitations impacts their ability to appropriately compare data to predictions or propose causes for discrepancies, and their ability to judge the quality of agreement between data and predictions impacts whether they revise their models or apparatus. We believe that it is possible to strike an effective balance between learning physics concepts and theories, and engaging in authentic modeling-based experimentation.} However, the dearth of research on teaching and learning in physics labs \cite{DBER2012} makes it difficult to know which instructional practices work well for particular learning goals, physics domains, and student populations.

In ongoing work, we are using the Modeling Framework to inform the design of standardized and scalable assessments of students' experimental modeling abilities \cite{Dounas-Frazer2018arXiv}. Interviews with instructors point to the need for process-based instruments that assess students' competence with multiple subtasks and the iterative nature of modeling, as well as their rationale for choosing one modeling pathway over another (e.g., deciding to revise the apparatus rather than collect more data). We aim to develop {instruments that are} compatible with a recent model for centralized data collection and large-scale deployment of research-based assessments~\cite{Wilcox2016b}. Doing so will enable us to identify individual courses or types of courses that are successfully improving students' experimental modeling abilities, which in turn will pave the way for further research on effective {modeling-oriented} teaching practices in physics labs. Ultimately, we hope that this work will create more opportunities for students to authentically engage in experimentation during their undergraduate physics education.

\ack
{We thank Benjamin Pollard and Laura R\'ios for useful discussions about the ideas presented here.} This material is based upon work supported by the National Science Foundation under Grant Nos.\ DUE-1726045 and PHY-1734006.

\section*{References}
\bibliography{modeling_database_180425}
\bibliographystyle{MYjphysicsB}

\end{document}